\documentclass[
aps,
prb,
twocolumn,
amsmath,
amssymb,
floatfix
]{revtex4-2}

\usepackage{booktabs}
\usepackage{graphicx}
\usepackage{tabularx}

\usepackage[separate-uncertainty=true,multi-part-units=single]{siunitx}
\usepackage{nicefrac}
\usepackage{hyperref}
\hypersetup{colorlinks=true, breaklinks=true, citecolor=blue, linkcolor=blue, menucolor=blue, urlcolor=blue}

\usepackage[dvipsnames]{xcolor}
\usepackage[normalem]{ulem} 
\DeclareSIUnit{\ML}{\ensuremath{ML}}

\newcommand{\figref}[1]{Fig.\,\ref{#1}}
\newcommand{\smfigref}[1]{SM-Fig.\,\ref{#1}}
\newcommand{\subfigref}[2]{Fig.\,\ref{#1}\,(#2)}
\newcommand{\tite}{TiTe\textsubscript{2}}
\newcommand{\rectcell}{$(5\times \sqrt 3 )_\textrm{rect}$}
\newcommand{\trt}{$\left(3\sqrt{3}\times3\sqrt{3}\right)\mathrm{R}30^{\circ}$}
\newcommand{\stmparams}[2]{{$U=\qty{#1}{\V}$, $I=\qty{#2}{\nA}$}}
\newcommand{\vipleed}{ViPErLEED}
\newcommand{\qfour}{quasi-$(4\times4)$}
\newcommand{\fourbyfour}{$(4\times4)$}

\begin{document}
	

\title{Growth and crystallographic structure of TiTe\textsubscript{2} on Au(111): \\From sub-monolayer structures to single- and multi-layer films}

\author{Andreas Raabgrund}

\author{Tilman Ki\ss linger}

\author{Alexander Wegerich}

\author{Lutz Hammer}

\author{M. Alexander Schneider}\email{alexander.schneider@fau.de}
\affiliation{Solid State Physics, Friedrich-Alexander-Universit\"{a}t Erlangen-N\"{u}rnberg, Staudtstra{\ss}e 7, 91058 Erlangen, Germany}
	

\begin{abstract}
	We investigated the initial growth of \tite\ on Au(111) from sub-monolayer to multi-layer coverage by scanning tunneling microscopy (STM), low-energy electron diffraction intensity analysis (LEED-IV), and density functional theory (DFT).
	In the submonolayer regime we find a stable and well-ordered $(5\times\sqrt{3})_{\mathrm{rect}}$ superstructure consisting of separated TiTe$_2$ molecules, whereby the Ti atoms substitute Au atoms of the first substrate layer as proven by LEED-IV.
	By adding further Ti and Te in a 1:2 ratio and proper annealing dealloying sets in and a homogeneous 1T-\tite\ monolayer film on an unreconstructed substrate is formed. The resulting moiré structure is close to a $(4 \times 4)$ superstructure w.r.t.\ Au(111) and has a slightly expanded in-plane lattice parameter compared to the 1T-\tite\ bulk value. With further stoichiometric deposition, thicker 1T-\tite\ films grow. 
	Surprisingly, a five layer thick film exhibits an even larger lattice-parameter (\qty{1.5}{\percent} larger than the bulk value).
	All LEED-IV analyses are based on best-fit R-factors of $R \le 0.13$.
\end{abstract}
	

\maketitle

\section{Introduction}
In the wake of graphene also the research of other two-dimensional materials has attracted considerable interest. Among these, transition metal dichalcogenides (TMDCs) of the form MX\textsubscript{2} (M = metal, X = S, Se, Te) have raised particular attention mostly due to their topological properties but also from an application point of view. 
The semimetal \tite\ \cite{Claessen1996,Rossnagel2001} attracted attention recently due to reports of a charge-density wave (CDW) in the single-layer limit \cite{Chen2017}, its use as electrode in batteries \cite{Antonio2025} or as carrier of single-atom catalysts \cite{Wang2017,Jia2025}. 

Further tuning and understanding of material properties depend on precise understanding of the crystallographic structure.
Particularly in the single-layer limit TMDCs often exhibit phenomena distinct from their bulk counterparts (e.g.\ change of character of the band gap \cite{Manzeli2017} or appearance of a CDW \cite{Chen2017}). Additionally, also the interaction with the substrate can induce new properties. When TMDCs are grown on crystalline substrates strain may develop due to mismatch of lattice parameters of film and substrate.
It has been shown, e.g. for MoTe\textsubscript{2}, that a reversible 2H to 1T' transition can be induced by moderate tensile stress \cite{Song2015, Awate2023}.
By epitaxial growth on a substrate with a different in-plane lattice constant, which indirectly mediates strain, modification of the intrinsic band structure and the properties of single- and multi-layered TMDCs were reported \cite{Fragkos2019}.
We demonstrate here that, when grown on Au(111), \tite\ is not forced into the nearby commensurate lattice parameter while the lattice parameters of single- and multi-layer films deviate from the bulk value.

When growing a single TMDC monolayer on various substrates, the interface formation can play a crucial role in understanding the growth processes and determining the geometrical and electronic structure. Therefore, investigating the reactions between the surface of the selected substrate and sub-monolayer amounts of the atomic species of the target TMDC is an important first step. Au(111) is one of the preferred substrates and is already well-established for the growth of other TMDCs \cite{Yasuda2017, Groenborg2015, Dendzik2015, Sanders2016}.
With respect to the growth of \tite\ on Au(111), the structures induced by depositing each of the two elements separately were already studied \cite{Guan2018, Potapenko2009}.

The epitaxial growth of \tite\ on a Te buffer layer on Au(111) was claimed by \citet{Song2019}.
In contrast, by a concerted application of LEED-IV, STM and DFT we will show here that the mentioned surface structure is in fact a sub-monolayer Au(111)-\rectcell-Ti\textsubscript{2}Te\textsubscript{4} structure, where the notation \rectcell\ is equivalent to the superstructure matrix $\left(\begin{smallmatrix}1 & 2 \\ 5 & 0\end{smallmatrix}\right)$. 
Additionally, we report on two sub-monolayer structures for Ti and Te on Au(111), a defective \trt\ and a defect-rich $(3\times 3)$. These structures partially coexist with a \tite\ monolayer film which grows on Au(111) when the amount of Te and Ti is correctly chosen. Following a specific growth sequence, single- and multilayer 1T-\tite\ films grow epitaxially on Au(111). From the first layer onward, the film adopts its appropriate, relaxed in-plane lattice parameter, forming a \qfour\ superstructure of $(3\times 3)$ \tite\ units with a \qty{1}{\percent} mismatch. LEED-IV analyses confirm the 1T-stacking sequence of the \tite-layers.

The present finding of a simple TMDC growth on a bare metal substrate is by no means the general case though quite often postulated, see e.g.\ the review by \textcite{Lu2021}. When TMDCs are grown by molecular beam epitaxy (MBE), be it by reaction of the metal substrate itself with the chalcogenide atoms X (X = S, Se, Te) or the reaction of foreign metal atoms with X on the substrate's surface, it is likely that a majority of systems follow the principles of quasi van-der-Waals (vdW) epitaxy \cite{Koma1984, Mortelmans2021}, where the interaction between the substrate and the growing film is not of pure vdW type.
This entails the appearance of buffer layers \cite{Nakano2017, Wang2018} and/or surface reconstructions. An example of the latter is the alloying of the Pt(111) surface with Te, which induces complex surface reconstructions that ultimately enable quasi vdW-epitaxy of Pt\textsubscript{2}Te\textsubscript{2} and PtTe\textsubscript{2} \cite{Kisslinger2023}. It is therefore always essential to verify the postulated film structures by sound structural analyses as e.g.\ performed in Refs.\,\cite{Kisslinger2020, Kisslinger2021, Kisslinger2023} and also in this study.

\section{Methods}
The Au(111) surface was cleaned by at least two cycles of sputtering with Ne ions and annealing to $\qty{800}{\K}$. Subsequently, high-purity Te ($\qty{99.999}{\percent}$, lump, Alfa Aesar) placed in a graphite crucible was deposited either from a home-built Knudsen cell evaporator (heated to $T=\qty{590}{\kelvin}$) or from an electron-beam evaporator at a rate of $\approx \qty{0.1}{\ML\per\minute}$. The Te amount was precisely calibrated by LEED with the aid of the well-known Te-induced superstructures on Cu(111) \cite{Kisslinger2020, Kisslinger2021} or on Pt(111) \cite{Kisslinger2023}. All LEED data presented here were taken at a sample temperature of \qty{100}{\K}.
In a second step, high-purity Ti ($\qty{99.99}{\percent}$, rod, Alfa Aesar) was deposited from an electron-beam evaporator at a rate of $\approx \qty{0.33}{\ML\per\minute}$. The Ti evaporation rate was first calibrated by measuring the frequency shift of a quartz microbalance and after the analysis of the \rectcell\ by repeated preparation of this Ti-Te structure. 
The experiments were carried out in two different ultra-high vacuum (UHV) chambers which are both separated into two independently pumped sub-chambers. One is used for sample preparation and LEED measurements ($p\le10^{-10}\si{\milli\bar}$). The other sub-chamber (base pressure $p\le\qty{5e-11}{\milli\bar}$) hosts either a beetle-type room-temperature STM with a tungsten tip or a home-built low-temperature STM with a PtIr tip operating at $\qty{80}{\kelvin}$. In both STM setups, the bias voltage is applied to the sample, while the tip was virtually grounded via the I/V-converter.

Experimental LEED-IV data were recorded at normal incidence by a cooled 12-bit CCD camera for energies up to $\qty{500}{\eV}$ in steps of $\qty{0.5}{\eV}$ and stored for off-line evaluation. The standard procedure for evaluating LEED data is described in Ref.\,\cite{Kisslinger2023}.
We used the newly developed \mbox{\textsc{ViPErLEED}} package \cite{Kraushofer2025, Schmid2025} which provides a sophisticated tool for LEED-IV data acquisition and manages a modified and parallelized \mbox{\textsc{TensErLEED}} code \cite{Blum2001} for full-dynamical calculation of intensity spectra and parameter fitting.
The employed lattice parameter for Au(111) at $\qty{90}{\K}$ was set to $a_{Au111}=\qty{2.8756}{\AA}$ \cite{latticeparameterAu}
and the respective bulk vibrational amplitude to $\qty{0.094}{\AA}$, according to a Debye temperature of $\Theta_{\text{D}}=\qty{165}{\K}$ \cite{Kittel1980}. The agreement between model intensities and experimental IV-curves was quantified using Pendry's R-factor \cite{Pendry1980}. Because of the known improper treatment of spin-orbit coupling for heavy elements like Au in present LEED codes, we neglected electron energies below $\qty{100}{\eV}$ in the analysis as suggested by \citet{Materer1995} when the Au substrate was taken into account.

Density functional theory (DFT) calculations were done with the Vienna Ab-initio Simulation Package (VASP) \cite{Kresse1996} using the PBE-PAW general gradient approximation \cite{Perdew1996}.
The \qfour\ structure was simulated by commensurate $(4 \times 4)$ slabs with different local registries w.r.t.\ the underlying substrate. Both, the  \rectcell\ slab and the $(4 \times 4)$ slab were constructed with six Au(111) layers and the respective \tite\ overlayers, whereby the lowest two Au layers were kept fixed. The energy cutoff was set to $\qty{300}{\eV}$ and $\qty{250}{\eV}$, respectively.
Repeated slabs along the surface normal were separated by $\qty{20}{\AA}$ and k-space was sampled with a $4 \times 10 \times 1$ and $4 \times 4 \times 1$ $\Gamma$-centered k-point mesh, respectively. The structures were relaxed until forces were smaller than $\qty{0.1}{\eV\per\AA}$.

\section{Ordered Ti-Te-phases: experimental findings}
\label{sec:results}

\subsection{Sub-monolayer structures}
\begin{figure}
	\centering
	\includegraphics[width=0.99\columnwidth]{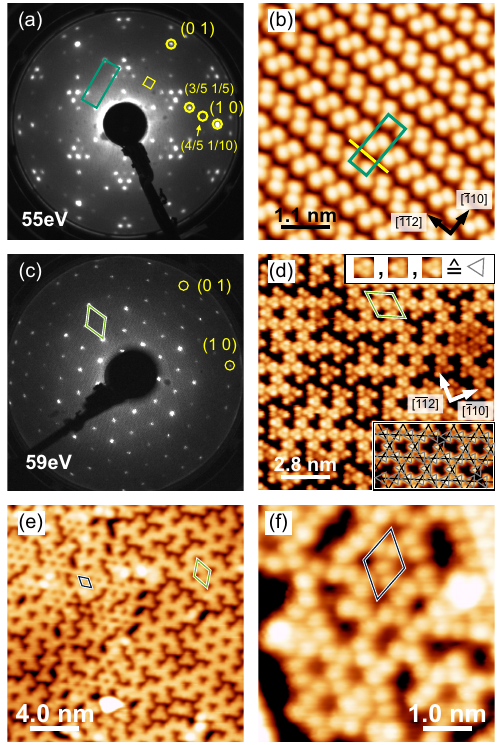}
	\caption{
	(a) After deposition of $\Theta_{Te}=\qty{0.40}{\ML}$ and $\Theta_{Ti}=\qty{0.20}{\ML}$ onto Au(111) and annealing to $T=\qty{470}{\K}$, the surface is covered by a well-ordered \rectcell\ superstructure (green rect.\,unit cell). 
	The missing spot at the (\nicefrac{1}{2} 0) position is marked by a yellow square in the LEED pattern. The circular yellow markers refer to the intensity spectra shown in \subfigref{Fig3_v1}{a}.
	(b) The close-up STM image shows the atomic basis of the \rectcell\ superstructure. 
	The yellow line indicates the glide plane.
	(c) After deposition of $\Theta_{Te}=\qty{0.44}{\ML}$ and $\Theta_{Ti}=\qty{0.11}{\ML}$ and post-annealing to \qty{420}{\K} LEED reveals a \trt\ pattern. 
	The first-order integer spots and the reciprocal superstructure cell are highlighted with yellow circles and with the green diamond, respectively. 
	(d) STM image revealing defect-rich \trt\ domains.
	The \trt\ unit cell is composed of building blocks formed by three protrusions in a triangular arrangement (upper inset) which are the building blocks of a kagome-like structure (see overlay). 
	(e) Exceeding the maximum Ti amount required for the \trt\ structure results in the formation of a defect-rich $(3\times 3)$ phase, see upper left area. 
	(f) Such an atomically resolved $(3\times 3)$ patch features three protrusions in the unit cell (annotated in blue).  STM parameters: (b) \stmparams{2.0}{0.2}; (d) \stmparams{0.1}{0.5}; (e) \stmparams{-1.0}{0.1}; (f) \stmparams{0.01}{2.0}
	}
	\label{Fig1_v1}
\end{figure}
The most stable superstructure at low Te and Ti coverage is the Au(111)-\rectcell-TiTe$_2$ at $\qty{0.4}{\ML}$ Te and $\qty{0.2}{\ML}$ Ti.
It may be reached by depositing the appropriate amounts of Te and Ti onto the Au(111) substrate at room temperature or below and subsequent annealing up to $\qty{470}{\kelvin}$.  At \qty{570}{\kelvin} the superstructure is completely absent in LEED which marks its thermal stability limit.
Alternatively, other structures discussed in this paper may be decomposed and thereby converted to the \rectcell\ structure by annealing to temperatures higher than their respective stability region.

With either preparation route, the Au(111) surface is uniformly covered with a \rectcell\ superstructure as revealed both by LEED and STM [see \subfigref{Fig1_v1}{a,b}]. LEED shows a clear and sharp \rectcell\ pattern (in three symmetry-equivalent domains) with a low background intensity. Accordingly, STM reveals huge mono-phase domains (sizes $\geq \qty{2000}{nm^2}$) that have an internal stripe-like appearance. 
The chains run along \{$\bar{1}\bar{1}2$\} directions of the crystal. 
The number of protrusion per unit cell matches the amount of offered Te atoms indicating that STM images just the Te atoms in line with the observation for other systems \cite{Kisslinger2020, Kisslinger2021, Uenzelmann2020}.
The systematic absence of the $(\nicefrac{1}{2}\;0)$ and equivalent spots in the LEED pattern [marked by a yellow square in Fig.\ref{Fig1_v1}(a)] reveals a glide plane within the unit cell normal to the $[\bar{1}10]$ direction. This is also apparent in the STM image [yellow line \subfigref{Fig1_v1}{b}], where neighboring chains are shifted mirror images of each other. 

The \rectcell\ phase turns out to be quite sensitive to Ti over- and underexposure. 
Based on the structural analysis of the \rectcell\ phase (see Section \ref{sec:5xsqrt3}), we can precisely quote the required Te and Ti amounts to form the following less-ordered Ti-Te phases on Au(111).
For example, underexposure of Ti and slight overexposure of Te with respect to the \rectcell\ phase leads, after annealing to \qty{420}{\K}, to the formation of a defective \trt\ superstructure containing \qty{0.44}{\ML} Te and \qty{0.11}{\ML} Ti (\subfigref{Fig1_v1}{c,d}).
Atomically resolved STM images reveal the \trt\ unit cell [green diamond in \subfigref{Fig1_v1}{d}]. It consists of three triangular building blocks with three protrusions each, one of which appears brighter than the other two, c.f.\ upper inset of \subfigref{Fig1_v1}{d}. Those buildings blocks are arranged on a kagome lattice, whereby the brighter protrusion always points towards the holes. 
With slightly increased Ti coverage at the surface, STM reveals a coexistence of defect-rich \trt\ and $(3\times 3)$ areas. Ordered $(3\times 3)$ patches exist that are a few unit cells in size, as shown in the upper left part of \subfigref{Fig1_v1}{e}. Atomically resolved STM images of such a patch reveal again a kagome-like arrangement, however, now build up by individual protrusions, cf.\ \subfigref{Fig1_v1}{f}. An analysis of area shares reveal that the $(3\times 3)$ phase hosts the same amount of Te (\qty{0.44}{\ML}) but \qty{0.22}{\ML} Ti. 
Thus, the $(3\times 3)$ structure has the very same stoichiometry as the \rectcell\ phase, but is just somewhat overexposed in both Ti and Te. 

\subsection{\tite\ monolayer film}
\label{sec:monolayer}

\begin{figure}
\centering
\includegraphics[width=0.99\columnwidth]{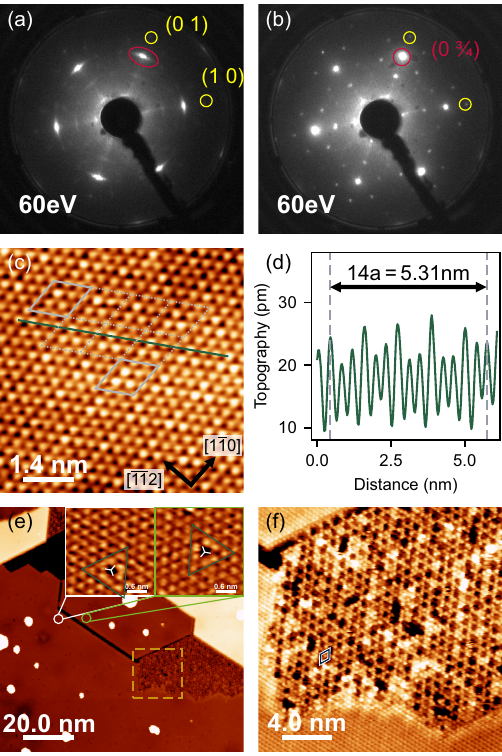}
\caption{
	(a) LEED pattern at $\qty{60}{\eV}$ after evaporation of $\Delta\Theta_{Te}=\qty{0.7}{\ML}$ and $\Delta\Theta_{Ti}=\qty{0.35}{\ML}$ onto the \rectcell\ structure and post-annealing to $\qty{620}{\K}$. Broadened $(0\;\nicefrac{3}{4})$ spots with centers still in direction of the substrate spots indicate only small misalignment angles of the \tite\ domains. (b) For the approach of stepwise preparation the LEED pattern ($\qty{60}{\eV}$) exhibits sharp spots. The structure is denoted as \qfour\ (for details see text). (c) Atomically resolved STM image of the \qfour\ reveals a moiré-like height modulation. The two solid gray $(3\times 3)$ meshes together with the dashed lines serve as a guide for observing the contrast variations across the surface. (d) STM height profile along the green line displayed in (c).	(e) The STM image for stepwise preparation reveals an almost perfectly grown layer as well as mirror domains. The insets show a zoom-in of equivalent defects in differently oriented domains. The orientation indicated by white three-pointed stars appears mirrored as expected for two different stacking sequences. (f) The STM close-up at the position of the dashed rectangle marked in (e) shows the defect-rich $(3\times 3)$ phase with the \tite\ layer. Imaging parameters: (c) \stmparams{0.01}{2.00}; (e) \stmparams{0.72}{0.21} (inset: \stmparams{0.03}{0.93}); (f) \stmparams{0.72}{0.21}
}
\label{Fig2_v1}
\end{figure}

If we recall that the lattice parameter of \tite\ is approximately $\nicefrac{4}{3}$ that of Au(111), we expect the formation of a single \tite\ layer with a $(4\times 4)$ superstructure, containing $(3\times 3)$ \tite\ unit cells.
The total Te and Ti coverage required is then $\Theta_{Te}=\nicefrac{9}{8}$\,ML and $\Theta_{Ti}=\nicefrac{9}{16}$\,ML.
For the preparation of such a monolayer \tite\ film we chose the well ordered \rectcell\ phases as a starting point and evaporated the remaining $\Delta\Theta_{Te}=\qty{0.72}{\ML}$ and $\Delta\Theta_{Ti}=\qty{0.36}{\ML}$ in a single step. After a successive annealing step to $T=\qty{620}{\K}$ indeed a $(4\times 4)$ structure develops.
However, prominent, but azimuthally broadened beams at $\nicefrac{3}{4}$ positions of the Au(111) reciprocal lattice vectors indicate the formation of an ordered \tite\ film with domains at a continuous distribution of angles of $\pm 6^\circ$ with respect to the Au(111) lattice, see \subfigref{Fig2_v1}{a}.
It is also noteworthy that the LEED pattern in this case is (almost) sixfold symmetrical. 

Strikingly, performing a stepwise preparation, i.e. offering a maximum of $\Delta\Theta_{Te}=\qty{0.2}{\ML}$ and $\Delta\Theta_{Ti}=\qty{0.1}{\ML}$ and annealing to $\qty{470}{\K}$ in steps until the required amounts of coverage are reached, leads after a final post-annealing to \qty{640}{\K} to a perfectly ordered $(4\times 4)$ LEED pattern with dominant ($0\;\nicefrac{3}{4}$) LEED spots without any further signs of rotational disorder, see \subfigref{Fig2_v1}{b}.
Furthermore, the LEED pattern has now a clear threefold symmetry. IV curves of the $(\nicefrac{3}{4}\;0)$ and $(0\;\nicefrac{3}{4})$ beams underscoring the three- and sixfold character in LEED are shown in \smfigref{Fig7:leed_symmetries}. 

Atomically resolved STM images [see \subfigref{Fig2_v1}{c}] reveal, at least at first glance, a $(3\times 3)$ periodic arrangement of hexagonally ordered protrusions, corresponding to a  $(4\times 4)$ superstructure with respect to Au(111). Because of the locally varying arrangement of 3 \tite\ per 4 Au units at the interface, the Te atoms in the uppermost layer of the film also exhibit slightly different apparent heights, either due to electronic or pure strain interactions. 
However, on closer inspection, one can see that the apparent height variation of these protrusions is not strictly periodic, but varies across the surface.
The gray $(3\times 3)$ mesh in \subfigref{Fig2_v1}{c} is intended to serve as a visual guide for identifying these variations. 
As an example, the top left solid mesh contains one bright atom, while the solid mesh in the middle contains three bright atoms. 
This is a clear indication that the structure is not a commensurate but a moiré phase. 
This finding also underlines that the growing \tite\ layer does not exactly adapt to the periodicity of the Au(111) surface. The analyzed spacing between protrusions of $a=\qty{3.81(6)}{\AA}$, measured by many LT-STM line profiles as the one in \subfigref{Fig2_v1}{d}, similarly points to an in-plane distance about 1\% smaller than $4/3\cdot a_{Au(111)}=\qty{3.834}{\AA}$ at \qty{100}{\K}.
Such a 1\% difference in lattice parameter can also be retrieved from a precise analysis of the \nicefrac{3}{4} order spot position, which, however, is also at the limit of resolution due to pixel size and spot width.
All in all, we have in fact an incommensurate or moiré superstructure referred to in the following as \qfour\ structure.

Large-scale STM images sometimes show wide domain boundaries. 
The STM pattern around two atomic scale defects on either side of the domain boundary [highlighted by a green triangle in the inset of \subfigref{Fig2_v1}{e}] are mirror symmetric to each other. This is an obvious hallmark for two mirror domains due to a stacking fault at the interface between film and substrate.
The analysis of the area shares reveals that the two domain orientations are distributed approximately at a ratio of 80:20 on the surface.
Later on, we will use LEED-IV to prove that the \qfour\ phase is indeed a single layer of 1T-\tite\ (see Section \ref{sec:StructAnalyses}). 
Large-scale STM images further reveal that the layer-by-layer growth is not perfect, i.e. with nominally one monolayer coverage, small islands of the second layer [small white dots in \subfigref{Fig2_v1}{e}] can already be found an an incompletely closed first \tite\ layer with $(3\times 3)$ patches in the spaces between. 
Consequently, weak intensities at third order positions also appear in the LEED pattern of \subfigref{Fig2_v1}{b}. 
Based on a careful estimation of the total amounts of Te and Ti bound in the \qfour\ and defect-rich $(3\times 3)$ regions according to their respective surface areas, and a comparison with the deposited amounts, it can be concluded that no further Te or Ti can be present at the \tite/Au(111) interface. It thus appears that in this particular system the \tite\ film indeed grows on a bare and unreconstructed substrate in contrast to other comparable systems like e.g. Pt(111) \cite{Kisslinger2023}. 

\citet{Chen2017} report of a $(2 \times 2)$ superstructure for a single-layer 1T-\tite, which they ascribe to a CDW forming below \qty{92}{\K}. 
While not-expected to be present in the LEED data at \qty{100}{\K}, our STM data taken at \qty{80}{\K} do not show any signs of such an electronically driven distortion. 
This may point to a suppression of the CDW due to the presence of the metal substrate \cite{Lin2020TiTe}.

\subsection{Multilayer film}
\label{sec:multilayer}
\begin{figure}
\centering
\includegraphics[width=0.99\columnwidth]{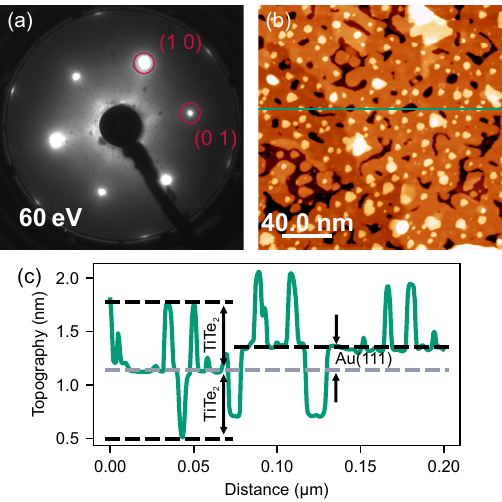}
\caption{
	(a) LEED pattern taken at \qty{60}{\eV} of a nominally 4.5 layer thick 1T-\tite\ film on Au(111). Due to the thickness of the film the \qfour\ superstructure w.r.t.\ Au(111) is no more visible and only the integer spots of the growing film remain. (b) Large-scale RT-STM image revealing a flat and homogeneously grown film. The green line annotates the position of the line profile displayed in (c). The observed step heights fit to either \tite\ and Au(111) layer distances, respectively. Imaging parameters: (b) \stmparams{2.05}{0.36} 
}
\label{Fig_multilayer_v1}
\end{figure}

These fully epitaxial single-layer 1T-\tite\ films are the starting point for a multilayer-film growth. The multilayer film shown in \figref{Fig_multilayer_v1} is grown with Te excess to avoid dissolution of Ti into the bulk due to annealing temperatures up to \qty{650}{\K}.
In total, $\qty{2.5}{\ML}$ Ti and $\qty{8.5}{\ML}$ Te are evaporated onto the Au(111) surface. 
Due to the Te excess, which desorbs during the annealing step, the final film thickness is defined by the offered Ti amount resulting in about 4.5 layers of \tite. 
The grown film exhibits a $(1\times 1)$ pattern (w.r.t.\ \tite) with no superstructure spots and threefold symmetry [see \subfigref{Fig_multilayer_v1}{a}]. Large-scale STM images reveal the growth of a homogeneous and flat film as displayed in \subfigref{Fig_multilayer_v1}{b}. 
\subfigref{Fig_multilayer_v1}{c} shows a line profile along the green line annotated in the STM image. The measured step heights correspond to the layer distances of either Au(111) or \tite, or linear combinations of both. The multilayer films are stable up to a temperature of about \qty{670}{\K}. 
Annealing to higher temperatures causes a successive reduction of the film thickness, subsequently leading to the reappearance of the \qfour\ and eventually the \rectcell\ phase in the LEED pattern.

\section{Structural Analyses}
\label{sec:StructAnalyses}
In this section we will focus on the crystallographic analyses for the \rectcell\ submonolayer phase as well as the single- and multilayer films.

For the \trt\ structure, there is so far no satisfying fit result. All tested models are revoked by a LEED-IV fit with Pendry R factors larger than $R=0.8$. 
These large R factors indicate wrong underlying structural ideas or the influence of coexisting phases. The latter could only be solved if at least one of the phases could be prepared as monophase and structurally determined which is not the case, unfortunately.

\subsection{Structure of the \rectcell\ phase}
\label{sec:5xsqrt3}

\begin{figure}
	\centering
	\includegraphics[width=0.99\columnwidth]{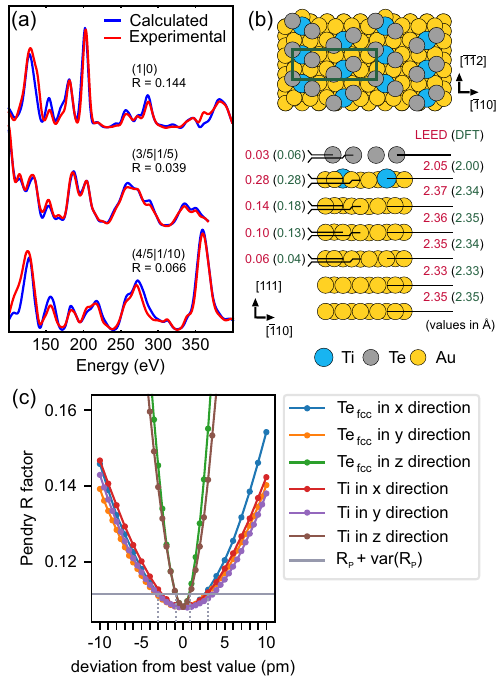}
	\caption{
		(a) Three exemplary intensity spectra (out of 71 \cite{seesm}) and comparison with the calculated bestfit spectra underline the quality of the fit. The positions of the displayed beams are marked in \subfigref{Fig1_v1}{a}.
		(b) Schematic ball model of the LEED-IV best-fit structure in top and side view with a Pendry R factor of $R + \mathrm{var}(R) = 0.106+0.006$  reveals incorporated Ti atoms in the topmost Au(111) layer and two Te atoms next to them residing approximately in hollow positions. The \rectcell\ unit cell is indicated in green. The side view (vertical distances exaggerated) visualizes the $\qty{28}{\pm}$ buckling of the chemically mixed top-layer and vertical relaxations down to the fourth Au(111) layer. The corresponding results of the DFT calculation (in brackets) match the LEED results within few picometers. 
		(c) Exemplary error curves for the (x,y,z) positions of the Te atom in nearly fcc position and the incorporated Ti atom resulting from the LEED-IV analysis. Vertical lines indicate the error margins. The full set of error curves is provided in the SM \cite{seesm}.
	}
	\label{Fig3_v1}
\end{figure}
The detailed crystallographic structure of the \rectcell\ phase is revealed by a concerted application of LEED-IV, STM, and DFT. 
All tested structural models are based on the pre-information from atomically resolved STM images and the amount of Te and Ti offered, which corresponds to 4 Te and 2 Ti atoms per \rectcell\ unit cell. 
A large number of models were relaxed using DFT and checked for compatibility with experiment by STM image simulation.
These relaxed structures were taken as starting point for the LEED-IV fit that is based on 71 independent experimental beams [example spectra are displayed in \subfigref{Fig3_v1}{a}] with a cumulated energy width of $\Delta E=\qty{13321}{\eV}$ ($\qty{1745}{\eV} / \qty{11576}{\eV}$ for integer/fractional order spots).
For the final best-fit model a total of $N=67$ fit parameters were optimized (61 structural, 3 vibrational, and 3 non-structural). Due to the huge data basis available even such a large number of free parameters could be safely determined by our LEED-IV analysis (redundancy factor $\rho=\nicefrac{\Delta E}{4NV_{0i}}=9.6$). 
The by far best-fit model with an excellent overall Pendry R factor of $R = 0.106$ provides a high-quality fit to the experimental data [see \subfigref{Fig3_v1}{a}]. The structural model consists of a row-like arrangement of single \tite\ ``molecules'' with every two of them within the \rectcell\ unit cell [see \subfigref{Fig3_v1}{b}]. 
The Ti atoms asume the position of a regular Au(111) surface atom but buckle out of the surface by $\qty{28}{\pm}$.
The Te atoms, on the other hand, assume approximately hollow adsorption sites next to the Ti atoms (one fcc and one hcp site each). 
Significant local relaxations, i.e. vertical bucklings and/or lateral displacements of the order of $\qty{0.1}{\AA}$, are found down to the fourth Au layer.
The maximum buckling amplitudes and average interlayer spacings, which were determined independently by LEED-IV and DFT, agree well within a few picometers, cf.\ \subfigref{Fig3_v1}{b}.  

An analysis of the variation of the R factor with parameter deviations from the best-fit values (''error curves'', for details see the SM Section \ref{sm:methods}) is shown in \subfigref{Fig3_v1}{c}. 
The position of the Te and Ti atoms can be determined in the unit cell to within $\pm\qty{1}{\pm}$  in the vertical direction and $\pm\qty{3}{\pm}$ laterally, underscoring the outstanding precision of our LEED-IV analysis.  
The whole set of error curves is provided in the SM \cite{seesm}.

\subsection{Structure of the monolayer \tite\ film}

\begin{figure}
\centering
\includegraphics[width=0.99\columnwidth]{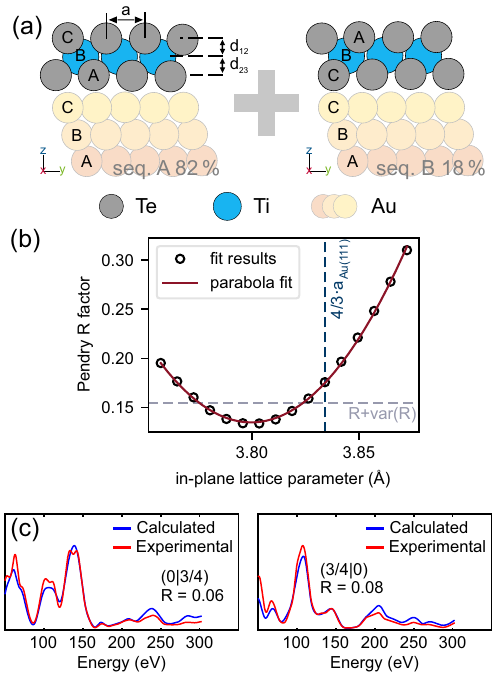}
\caption{
	(a) The LEED-IV domain fit approach finds a preference of \qty{82(9)}{\percent} for stacking sequence A while the rest is stacking sequence B.  
	(b) The optimization of the lateral lattice parameter results in a minimum Pendry R factor of $R+\mathrm{var}(R)=0.134+0.021$ 
	for a value of $\qty{3.800(24)}{\AA}$.  The commensurate lattice constant $\nicefrac{4}{3}\cdot a_{\mathrm{Au(111)}}$ is clearly outside the variance interval of the R factor. The best-fit values for the parameters specified in (a) are listed in Tab.\,\ref{tab1}. (c) The two first order beams of the growing \tite\ film are well represented by the calculated best-fit spectra and underline the quality of the fit. The full set of spectra is provided in the SM \cite{seesm}. 
}
\label{Fig4_v1}
\end{figure}%

\begin{table}%
	\begin{tabular}{rccc}
		\hline
		\rule[1pt]{0pt}{12pt} \quad parameter \quad & \quad single layer \quad & \quad multi layer \quad & bulk \cite{Arnaud1981} \\ \hline
		\rule[3pt]{0pt}{12pt} a \qquad& \qty{3.800(24:24)}{} & \qty{3.835(20:20)}{} & \num{3.777(3)} \\
		\rule[0pt]{0pt}{12pt} c \qquad& --- & \qty{6.448(36:31)}{} & \num{6.498(6)} \\
		\rule[0pt]{0pt}{12pt} $d_{12}$ \qquad& \qty{1.663(25:26)}{} & \qty{1.664(23:24)}{} & \num{1.708(2)} \\
		\rule[0pt]{0pt}{12pt} $d_{23}$ \qquad& \qty{1.695(19:17)}{} & \qty{1.708(23:24)}{} & \num{1.708(2)} \\
		\rule[0pt]{0pt}{12pt} $d_{\mathrm{vdw}1}$ \qquad& --- & \qty{3.042(16:14)}{} & \num{3.083(3)} \\
		\rule[0pt]{0pt}{12pt} $d_{45}$ \qquad& --- & \qty{1.696(34:28)}{} & \num{1.708(2)} \\
		\rule[0pt]{0pt}{12pt} $d_{56}$ \qquad& --- & \qty{1.656(48:59)}{} & \num{1.708(2)} \\ 
		\rule[0pt]{0pt}{12pt} $d_{\mathrm{vdw}2}$ \qquad& --- & \qty{3.042(32:38)}{} & \num{3.083(3)} \\ \hline
	\end{tabular}
	\caption{Best-fit lattice parameter and layer distances for single and multi-layer \tite\ films on Au(111) resulting from the LEED-IV analysis. The values are given in angstrom (\AA). The errors are calculated full-dynamically, for details see SM \cite{seesm}.}
	\label{tab1}
\end{table}%

The crystallographic analysis of the monolayer \tite\ film by LEED-IV requires an advanced strategy since the film is, as shown before, not a commensurate \fourbyfour\ superstructure but a moiré phase, i.e. no periodic unit cell can be defined. 
To discriminate between possible film structures nevertheless, we neglected the scattering contribution from the Au(111) substrate  and the local atomic displacements due to the moiré. This was achieved by using only the $(0\;\nicefrac{3}{4})$ and $(\nicefrac{3}{4}\;0)$ beams and their higher orders, i.e.\ treating the film as quasi free-standing with $1 \times 1$ periodicity.
This data was compared to calculation of with free-standing single layers of 1T-\tite, 2H-\tite, a TiTe bilayer, and a Ti\textsubscript{2}Te\textsubscript{2} layer. The latter two we tested to cross-check our Ti and Te amounts. 
An overview on these structural models, the fit results, and overall R factors is compiled in \smfigref{Fig4_v0}.
In the course of the fit we further optimized the area ratio of the two mirror domains as well as the lateral lattice parameter.
More detailed information on this procedure and how to achieve this with the \vipleed-package \cite{Kraushofer2025} is given in the SM Section \ref{sm:methods} \cite{seesm}.

The best-fit result undoubtedly proves the growth of a 1T-\tite\ film with a Pendry R factor of $R+\mathrm{var}(R)=0.134+0.021$, which is a surprisingly good result given the simplifications that have been made. 
The optimization of the ratio of the two mirror domains (stacking sequence A and B, cf.\ \subfigref{Fig4_v1}{a}) resulted in a ratio of $\qty{82}{\percent}:\qty{18}{\percent}$ with an error of $\pm\qty{9}{\percent}$, thereby quantitatively confirming the STM estimate. 
Finally, the fit reveals a slight lateral contraction of the film with respect to a commensurate \fourbyfour\ structure, as can be seen from the dependence of the R factor on the in-plane lattice parameter displayed in \subfigref{Fig4_v1}{b}. 
A parabolic fit finds a minimum at $\qty{3.800(24)}{\AA}$.
As before the error margin is obtained from the crossing points of the R-factor curve with the $R+\mathrm{var}(R)=0.155$ level \cite{Pendry1980}. 

A lattice parameter commensurate to the substrate at \qty{100}{\K} ($\nicefrac{4}{3}\ a_\textrm{Au(111)}=\qty{3.834}{\AA}$) is clearly outside the error margins of the R factor (the range below the variance level), see \subfigref{Fig4_v1}{b}. 
This finding clearly corroborates the estimates of the lattice parameter drawn from STM height profiles and LEED spot positions in Sec.\,\ref{sec:monolayer}, which, however, were  both at their resolution limit.  
The reported lattice parameter of a 1T-\tite\ bulk crystal is $a=\qty{3.777(3)}{\AA}$ (for room temperature) \cite{Arnaud1981, Riekel1978}, which is at the lower edge of our error bar. 
This means that the single-layer 1T-\tite\ grown on Au(111), which, as a moiré phase, should be laterally relaxed, does not completely adopt a bulk-like lattice parameter, but rather a slightly larger value, possibly due to interaction with the surface. 
In line with this lateral expansion is the finding of slightly compressed vertical layer distances for this monolayer film, see Tab.\,\ref{tab1}, so that the resulting Ti-Te bond lengths [\qty{2.76(5)}{\AA} and \qty{2.77(5)}{\AA}] were both at the resolution limit. This lateral expansion comes along with slightly compressed vertical layer distances (see Tab.\,\ref{tab1}). The outermost layer distance $d_{12}$ is found to be even smaller by \qty{3}{\pm} which is attributed to a surface relaxation as it also occurs in the multilayer \tite\ film, see below. 
Assuming normal thermal expansion, the difference cannot be attributed to the lower temperature of our experiment (\qtylist{100}{\K}) but comes from the (screening) interaction with the substrate.

\subsection{Structure of the multilayer \tite\ films}

\begin{figure}%
\centering
\includegraphics[width=0.99\columnwidth]{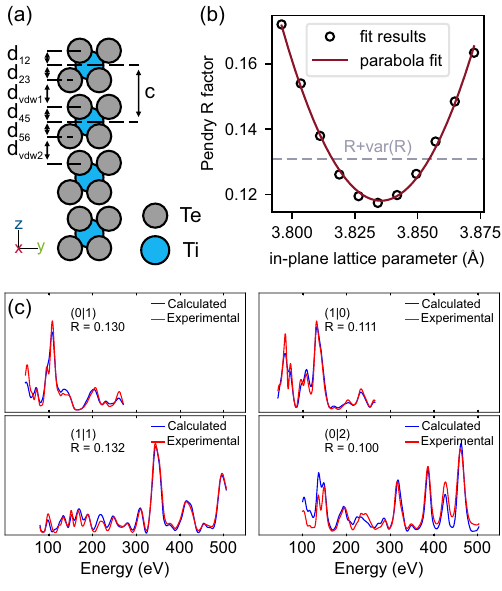}
\caption{
	(a) The LEED-IV analysis results in a bulk-like 1T-\tite\ film with a Pendry R factor of $R+\mathrm{var}(R)=0.120+0.014$. The best-fit values of the annotated parameters are listed in Tab.\,\ref{tab1}. (b) According to LEED-IV the lattice parameter of the film is $a=\qty{3.835(20)}{\AA}$. (c) Four exemplary intensity spectra and comparison with the calculated intensities from the best-fit model underscore the quality of the fit. The full set of spectra is provided in the SM \cite{seesm}.
}
\label{Fig6_v1}
\end{figure}%

For the nominally 4.5 layer thick film discussed in Sec.\,\ref{sec:multilayer} we also performed a LEED-IV structure analysis. As this film is sufficiently thick, so that the Au(111) substrate does no more contribute due to the limited penetration depth of the electrons, we simply treated it as a $1 \times 1$ \tite\ bulk structure. Similar to the monolayer film a domain fit approach was used to take the mirrored stacking sequence into account. 
The most relevant results of the fit with a Pendry R factor of $R+\mathrm{var}(R)=0.120+0.014$ are shown in \figref{Fig6_v1} and optimized values for structural parameters are compiled in Tab.\,\ref{tab1}.  For more details, see Supplement Section \ref{sm:subsec_multi}.

The resulting share of \qty{18(11:8)}{\percent} stacking fault domains remains unchanged with respect to the monolayer film indicating that the film growth just continues the stacking direction defined by the first \tite\ layer. 
As evident from Tab.\,\ref{tab1}, the topmost Te atom is relaxed inwards by about \qty{4}{\pm} which we attribute to surface relaxation.
Consequently, we define the c parameter, i.e. the distance between the topmost trilayer, by the difference in z values of the two topmost Ti atoms, see \subfigref{Fig6_v1}{a}.
The found value of $c=\qty{6.448(36:29)}{\AA}$ is approximately \qty{1}{\percent} smaller than the bulk value ($c=\qty{6.498(6)}{\AA}$ \cite{Arnaud1981}).
The outermost two vdW gaps significantly undercut the bulk value by \qty{0.04}{\AA}, c.f.\ Tab.\,\ref{tab1}. These reduced vdW gaps almost completely account for the difference in the c parameter with respect to the bulk value.
 
The optimization of the lateral lattice parameter resulted surprisingly in $a=\qty{3.835(20)}{\AA}$, a value larger than that of the monolayer film.
It is thus even farther away (approx. \qty{1.5}{\percent}) from the reported bulk value \cite{Arnaud1981}.
An increased in-plane lattice parameter compared to the value for a single-layer was also reported by \citet{Fragkos2019} for a \qty{50}{\ML} \tite\ film grown by MBE on InAs(111). 

\section{Discussion}
\label{sec:discussion}

Our investigation of the initial growth of \tite\ on Au(111) established a \rectcell\ surface phase with 1:2 Ti:Te ratio but only \qty{0.2}{\ML} Ti content. A surface structure of the same supercell was reported by \citet{Song2019} and interpreted as a \tite\ monolayer with Te buffer layer at the interface to the Au(111) substrate, however, without evidence from a crystallographic analysis.

We carefully analysed the similarity between STM images and dependence of surface structures on the amount of Ti and Te deposited. We come to the conclusion that \citet{Song2019} indeed prepared the same \rectcell\ phase as we analysed here by LEED-IV. The misinterpretation in Ref.\,\cite{Song2019} is possible due to an experimental misjudgement of the amounts of Ti and Te on the surface.  Particularly when working at higher substrate temperatures, the desorption of Te has to be taken into account as shown by \citet{Guan2018}.

While in the \rectcell\ phase the Ti atoms are found to replace first layer Au atoms, the \tite\ film grown by depositing more material onto this structure has neither Ti nor Te atoms in the interfacing Au layer.
Thus, dealloying takes place when more than \qty{0.2}{\ML} of Ti are present at the surface. 
A detailed analysis of the interface structure by LEED-IV is very complicated due to the moiré structure, which leads to a continuously shifting local registry between the film and substrate. 
How to derive reliable crystallographic information from LEED in such a situation will be subject of a separate forthcoming publication. 

It is surprising that despite the direct interaction with the Au(111) substrate the incommensurate (and thus fully relaxed) lattice parameter of a single \tite\ layer on Au(111) is only \qty{0.6}{\percent} larger than the bulk value, while a two to five layer thick film deviates even stronger with an \qty{1.5}{\percent} expansion with respect to \tite\ bulk value at room temperature (see Tab.\,\ref{tab1}). These expansions get even larger if the unknown lattice parameter of \tite\ at \qty{100}{\K} were considered. 
Therefore, the films are too thin to represent bulk properties. \citet{Fragkos2019} report that this is also the case for an even \qty{50}{\ML} thick \tite\ film on InAs(111). The reason for this remains unclear. 

In order to better understand why \qty{20}{\percent} of the \tite\ single- and multi-layer films grow with a stacking fault at the interface, DFT calculations for commensurate $(4\times 4)$ slabs were performed. 
The total energies derived from DFT for the two stacking sequences in various registries to the substrate are all degenerate within \qty{5}{meV} and therefore indicate neither a preferred stacking direction nor interface registry of the single-layer 1T-\tite\ film on Au(111). A similar result is reported, e.g., for 2H-TaS\textsubscript{2} on Au(111) \cite{Silva2021}.
Therefore, we suggest that the imbalanced ratio of the two stacking domains is triggered by defects at the Au(111) surface, most likely by step edges. A tiny misorientation of the surface that is unavoidable in practice could cause the deviation from the 50:50 ratio of the domains.
The idea that step edges determine the stacking direction is also supported by the described procedure, which is necessary to produce a well-ordered monolayer film without rotational disorder, cf.\ Sec.\,\ref{sec:monolayer}. Only if Ti and Te are deposited in small portions at the surface they have a good chance of reaching a step edge (or an already nucleated \tite\ island there) during the subsequent annealing step, where one stacking direction is favored. The larger the deposited doses, the higher the probability for on-terrace nucleation without preferred stacking. The same holds for lower annealing temperatures and thus shorter diffusion lengths, where indeed a more balanced domain ratio is found, as can be inferred from more similar spectra for e.g.\ $(\nicefrac{3}{4}\;0)$ and $(0\;\nicefrac{3}{4})$ beams, see \smfigref{Fig7:leed_symmetries}. Hence, the ratio of ``A'' and ``B'' domains appears to depend strongly on the annealing temperature, the deposition rate, and the step density (i.e. miscut angle and direction) at the surface. Conversely, all these variables can be used to specifically control the desired growth behavior. 

\section{Summary}
\label{sec:summary}
In summary, we present here a comprehensive analysis of the \tite\ growth on Au(111) based on LEED-IV, DFT, and STM results.
An epitaxial single-layer growth of \tite\ on Au(111) is achieved by a sequential growth recipe starting from a Au(111)-\rectcell\ surface structure with two \tite\ molecules in the unit cell arranged in linear rows, with the Ti atoms embedded in the first substrate layer. 
Our findings clearly disprove that the \rectcell\ structure is a \tite\ monolayer on Au(111) with a Te buffer layer as reported by \citet{Song2019}.
In fact, such a \tite\ monolayer is an almost commensurate, \qfour\ moiré structure of the 1T polymorph with an in-plane lattice parameter close to that of bulk \tite. 
There is neither Te nor Ti in the interface, i.e. the \tite\ film is in direct contact with the plain and unreconstructed Au(111) surface.
On the first layer further layers of 1T-\tite\ may be grown. 
By LEED-IV we prove that these multilayer films are laterally slightly more expanded than the monolayer film and additionally have a slightly smaller vdW gap than the bulk crystal.

\subsection{Acknowledgments}
We gratefully acknowledge support from the Deutsche Forschungsgemeinschaft (DFG), project 497265814. The ASE python package was partially used to create atomic slab models and plot the results \cite{Larsen2017}.

\subsection{Supplemental Material}
The supplement \cite{seesm} offers detailed information on the performed LEED-IV analyses as well as additional data referenced in the main text. Additionally, it provides a description of the electronic files included in the supplement. It also includes the additional reference Ref.\,\cite{Rundgren2003}.

\bibliography{literatur_TiTe2}
\pagebreak
\onecolumngrid
\appendix

\section{Methods}
\label{sm:methods}

\subsection{LEED-IV fits neglecting the substrate}

To fit a free-standing (TMDC) layer with the \vipleed\ code conflicts with the mandatory requirement that ``bulk'' layers have to be present in the model that normally would represent the substrate layers unaffected by the surface overlayer. 
To resolve this conflict, a model is defined which contains a Au bulk but where
the scattering strength of the Au substrate atoms is forced to practically zero by setting the occupation probability of the corresponding atomic sites to 0.001 (occupation parameter in the VIBROCC file) \cite{vipleed-man-files}.
In that way the distance between the substrate-defining Au sites becomes also irrelevant such that a $1\times 1$ unit cell of the \tite\ film can be used by adapting the lattice constant of the ``substrate'' to that of the film. 

The two mirror-symmetric stacking sequences identified by the STM images would require in principle a two-domain fit. 
However, for the first identification of the correct structural models among those shown in \figref{Fig4_v0}, a simplified approach was followed.
Rather than the full two-domain fit, an artificial 50:50 domain ratio was enforced by averging the respective experimental beams [e.g.\ $(\nicefrac{3}{4}\;0)$ and $(0\;\nicefrac{3}{4})$] in order to yield sixfold symmetry.
Performing the fit with an also artificial hcp Au bulk (with close to zero scattering strength), which would just produce such two equivalent mirror domains, leads to automatic consideration of symmetry equivalent domains within the ViPErLEED code. 
Given this approach, the 1T-\tite\ model leads (after fitting layer positions and vibrational amplitudes) to an R factor of $R=0.24$. 
The other models, namely the 2H-\tite, the TiTe bilayer, and the Ti\textsubscript{2}Te\textsubscript{2} layer, yield R factors that are twice as large. Thus, these models can be excluded already at this stage of the analysis with confidence.

In the next step, the fit procedure for the 1T-\tite\ layer was refined. The minority domain with stacking fault at the interface as found by STM was now taken into account by a \vipleed\ domain fit approach. The stacking fault domain was initialized as a true mirror image of the \tite\ layers of the stack domain, see main text \subfigref{Fig4_v1}{a}. 
Unfortunately, at present the \vipleed\ code (version $0.12.2$) does not allow to couple fit parameters between different domains. 
As a workaround, the majority ''stack`` domain solely was optimized and the atom positions within the film were mirrored after each fit cycle. This procedure ensured a lower number of fit parameters and thus a more stable fit as well as a higher redundancy. 

\subsection{Error calculation of fit parameters for a domain calculation}

In order to determine error magins for the fit parameters, a  single parameter is varied with all others are kept fixed at their best-fit value and corresponding IV-spectra are calculated. The R factor between calculated and experimental spectra is then plotted as function of deviation from the bestfit value of the parameter. 
The error margin of the parameter value is defined by the range where $R \leq R+\textrm{var}(R)$ with $\textrm{var}(R)= R \cdot \sqrt{\frac{8 V_{0i}}{\Delta E}}$ \cite{Pendry1980}.
The \vipleed\ package \cite{Kraushofer2025, Schmid2025} includes a tool, which calculates the errors using the tensor-LEED approach \cite{Blum2001, Rous1986, Rous1989}. Since this implemented error calculation can be used for single domains only, for a domain calculation a workaround is needed. 
If the statistical weight of the majority domain is dominating ($>\qty{80}{\percent}$) and its fit achieves a decent Pendry R factor below 0.2, the default error calculation for the majority domain is sufficient. In principle, this would work adequately with the multilayer \tite\ film on Au(111). The error curves, however, typically do not display a R factor minimum at their current best-fit position, as the structure has been optimized within a domain calculation. Therefore, we used a workaround here.

The full error curve calculation for the single- and multi-layer domain fit was achieved in the following way:
Full-dynamical reference calculations were executed for each domain for sequentially altered parameters (here: z positions and vibration amplitudes only). The calculated beams were incoherently superimposed according to the best-fit statistical weights of the domains. The Pendry R factor of each configuration with respect to the experimental data was calculated using the R factor routine of the \vipleed\ ImageJ plugin \cite{Schmid2025}. 
The resulting error curves for the single- and multi-layer \tite\ film are shown in \path{error_full-dynamic_singlelayer_tite2.pdf} and \path{error_full-dynamic_multilayer_tite2.pdf}, respectively. The derived layer spacings, along with their associated errors, are presented in Tab.\,\ref{tab1} of the main text. Note that the error for the two outermost vdW gaps was calculated while all other atomic distances were kept fixed.

A similar workaround for the optimization routines of the \vipleed\ package, such as those for the in-plane lattice parameter and the domain ratio, is used. 
Eventually, full-dynamical calculations for each domain and for each variation step are executed, incoherently superimposed and compared to experimental IV data.

\section{Ordered Ti-Te-phases: experimental findings}

\begin{figure*}[h]
	\centering
	\includegraphics[width=0.99\textwidth]{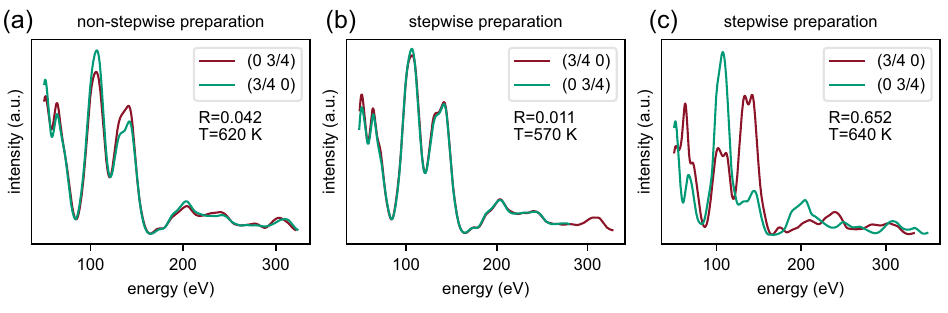}
	\caption{
		IV curves of the $(0\;\nicefrac{3}{4})$ and $(\nicefrac{3}{4}\;0)$ beams for (a) non-stepwise and (b, c) stepwise preparation (each averaged over the three symmetrically equivalent beams according to the p3m1 symmetry of the Au(111) substrate): The displayed Pendry R factors are used to quote the degree of similarity between the two beams. The given temperature is the post-annealing temperature. Utilizing similarly high post-annealing temperatures (\qty{620}{\K} and \qty{640}{\K}), the stepwise preparation method results in a threefold symmetry (for further details, refer to the main text), whereas the non-stepwise approach generates a sixfold symmetric LEED pattern. Lower post-annealing temperatures (such as \qty{570}{\K}) following the stepwise preparation also result in a sixfold symmetric LEED pattern. The IV curves displayed in (a) and (c) correspond to the LEED patterns shown in \subfigref{Fig2_v1}{a} and \subfigref{Fig2_v1}{b} of the main text, respectively.
	}
	\label{Fig7:leed_symmetries}
\end{figure*}%

\section{Details of the structural analyses}

\begin{figure*}[h]
	\centering
	\includegraphics[width=0.99\textwidth]{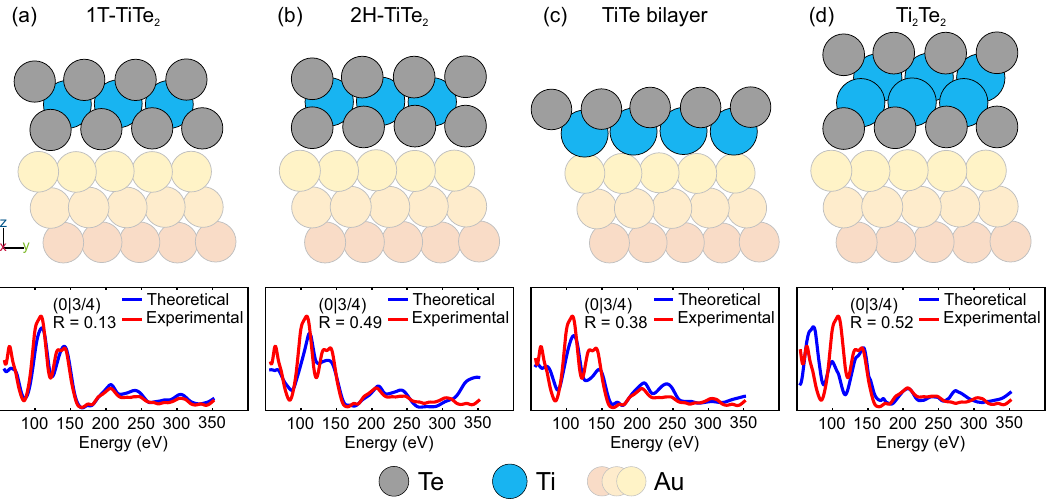}
	\caption{
		Tested structures by LEED-IV with neglected Au(111) substrate: (a) 1T-TiTe\textsubscript{2}, (b) 2H-TiTe\textsubscript{2}, (c) TiTe bilayer, and (d) Ti\textsubscript{2}Te\textsubscript{2}. As a representative beam the experimental and theoretical (0 3/4) beam are shown. The corresponding Pendry R factors are printed below each beam label. The overall Pendry R factors are (a) $R_{all}=0.24$, (b) $R_{all}=0.51$, (c) $R_{all}=0.55$, and (d) $R_{all}=0.54$ (for more details, see main text).
	}
	\label{Fig4_v0}
\end{figure*}%

The supplied archive \path{tite2_au111_supplement.zip} contains all data for each presented phase in a correspond sub-folder. 
These are the input and output files of the \vipleed-package \cite{Kraushofer2025}: the experimental LEED-IV data \path{EXPBEAMS.csv}, the bestfit structure \path{POSCAR_bestfit} as input file compatible with VASP \cite{Kresse1996}, the fitted vibrational amplitudes \path{VIBROCC_bestfit}, the graphical representation and comparison between calculated and experimental spectra \path{rfactor_plots_bestfit.pdf} and the R factor as function of deviation from the bestfit parameter value \path{error_*.pdf}.
In case of the \rectcell\ a numerical list of the errors \path{errors_5xsqrt3-rect.csv} and the DFT-relaxed structure \path{CONTCAR} are provided, additionally.
    
\label{sm:details}
\subsection{\rectcell\ fit}
Table \ref{tab:rect} presents the results of the structural analysis for the Au111-\rectcell-2(\tite) superstructure. The structural parameter values indicate deviations (in \AA) from the idealized Au(111) structure (including the two substituted Ti atoms). The position of the uppermost Te atoms is related to the precise hcp and fcc hollow sites and the height of the next Au layer. The vibrational amplitudes, expressed in \AA, are absolute values derived from the LEED-IV fit. Notably, the vibrational amplitude for Au bulk is set at $\qty{0.094}{\AA}$, corresponding to a Debye temperature of $\Theta_{\text{D}}=\qty{165}{\K}$ \cite{Kittel1980}.
Additionally, further non-structural parameters were optimized: the optical potential $V_{0i}$ and a small, rigid energy shift $V_{00}$, which accounts for the unknown $E\to\infty$ level of the energy dependent inner potential \cite{Rundgren2003}, and an effective angle of incidence $\Theta_{\mathrm{eff}}$ to model the conical shape of the incident electron beam. After optimization, the effective angle of incidence is determined to be $\Theta_{\mathrm{eff}} = \qty{0.33}{\degree}$ and the optical potential to $V_{0i}=\qty{5.2}{\eV}$. The total energy data basis amounts to $\Delta E = \qty{13321}{\eV}$, and in total, $N=67$ fit parameters are utilized. The redundancy of the fit is given by $\rho = \nicefrac{\Delta E}{4NV_{0i}} \approx 9.6$.

\begin{table}
	\begin{tabular}{llrrrrrrr}
		\toprule
		N & atom (label) & \multicolumn{2}{c}{$\Delta x$} & \multicolumn{2}{c}{$\Delta y$} & \multicolumn{2}{c}{$\Delta z$} & \multicolumn{1}{c}{$\Delta u$} \\
		& & \multicolumn{1}{c}{LEED} & \multicolumn{1}{c}{DFT} & \multicolumn{1}{c}{LEED} & \multicolumn{1}{c}{DFT} & \multicolumn{1}{c}{LEED} & \multicolumn{1}{c}{DFT} & \multicolumn{1}{c}{LEED} \\
		\midrule
		\rule[0pt]{0pt}{10pt}  3  & Te (Te\_def)   &  \num{-0.236(35:40)}  &  -0.200  &   \num{0.100(44:40)}  &  0.112  &  \num{-0.261(11:11)}  &   -0.354  &  \num{0.137(15:11)}  \\
		\rule[0pt]{0pt}{10pt}  5  & Te (Te\_def)    &   \num{-0.287(38:41)}  &  -0.242  &   \num{0.010(36:48)}   &  0.149   &   \num{-0.298(12:10)}  &   -0.409  &                      \\
		\rule[0pt]{0pt}{10pt}  1  & Ti (Ti\_def)       &   \num{0.021(37:37)}   &   0.028  &   \num{0.057(42:42)}   &  0.124   &   \num{0.190(12:11)}   &   0.114   &  \num{0.128(18:19)}  \\
		\rule[0pt]{0pt}{10pt}  7  & Au (Au\_surf)  &   \num{0.050(34:40)}   &   0.051  &   \num{0.116(54:48)}   &  0.206   &   \num{0.027(26:26)}   &   0.019   &  \num{0.121(15:15)}  \\
		\rule[0pt]{0pt}{10pt}  9  & Au (Au\_surf)   &   \num{-0.035(43:37)}  &  -0.072  &   \num{-0.010(49:45)}  &  0.045   &   \num{-0.014(23:24)}  &   -0.031  &                      \\
		\rule[0pt]{0pt}{10pt} 11 & Au (Au\_surf)   &   \num{-0.043(43:43)}  &  -0.045  &   \num{0.152(46:43)}   &  0.253   &   \num{-0.093(21:19)}  &   -0.169  &                      \\
		\rule[0pt]{0pt}{10pt} 13 & Au (Au\_surf)  &   \num{0.064(46:47)}   &   0.080  &   \num{-0.066(45:42)}  &  -0.003  &   \num{-0.010(17:15)}  &   -0.078  &                      \\
		\rule[0pt]{0pt}{10pt} 15 & Au (Au\_def)  &   \num{0.000(51:42)}   &  0.018  &   \num{0.010(48:50)}   &  0.075   &   \num{0.003(20:22)}   &   -0.048  &    0.094           \\
		\rule[0pt]{0pt}{10pt} 16 & Au (Au\_def)  &   \num{-0.020(45:37)}  &   -0.026  &   \num{-0.040(45:40)}  &  -0.007   &   \num{0.057(18:21)}   &   0.049  &                 \\
		\rule[0pt]{0pt}{10pt} 17 & Au (Au\_def)  &   \num{-0.010(45:43)}  &   -0.012  &   \num{0.040(42:41)}   &  0.113  &   \num{-0.075(14:16)}  &   -0.130   &                      \\
		\rule[0pt]{0pt}{10pt} 18 & Au (Au\_def)  &   \num{-0.010(33:34)}  &  -0.014  &   \num{0.010(48:50)}   &  0.067  &   \num{-0.015(21:23)}  &   -0.049   &                      \\
		\rule[0pt]{0pt}{10pt} 19 & Au (Au\_def)  &   \num{0.040(38:33)}   &   0.066  &   \num{0.010(53:62)}   &  0.042   &   \num{0.059(26:25)}   &   0.055  &                      \\
		\rule[0pt]{0pt}{10pt} 25 & Au (Au\_def)  &   \num{0.000(43:52)}   &   0.000  &   \num{0.000(81:85)}   &  0.031   &   \num{0.049(31:35)}   &   0.039  &                      \\
		\rule[0pt]{0pt}{10pt} 26 & Au (Au\_def)   &   \num{0.000(62:63)}   &  -0.008  &   \num{0.000(56:57)}   &  0.041   &   \num{0.003(25:23)}   &   -0.035  &                      \\
		\rule[0pt]{0pt}{10pt} 27 & Au (Au\_def)  &   \num{0.020(66:54)}   &  0.012  &   \num{-0.010(57:57)}  &  0.025   &   \num{0.011(25:24)}   &   -0.012  &                      \\
		\rule[0pt]{0pt}{10pt} 28 & Au (Au\_def)  &   \num{0.010(68:58)}   &   0.014  &   \num{0.010(56:56)}   &  0.053   &   \num{-0.050(26:29)}  &   -0.088  &                      \\
		\rule[0pt]{0pt}{10pt} 29 & Au (Au\_def)  &   \num{-0.010(48:50)}  &  -0.013  &   \num{-0.010(84:66)}  &  0.044   &   \num{-0.034(36:30)}  &   -0.062  &                      \\
		\rule[0pt]{0pt}{10pt} 35 & Au (Au\_def) &   0.000   &  -0.006  &   0.000   &  0.011  &   \num{0.006(27:28)}   &   -0.012  &                      \\
		\rule[0pt]{0pt}{10pt} 36 & Au (Au\_def)  &   0.000   &   0.005  &   0.000   &  -0.005  &   \num{0.025(46:47)}   &   -0.006  &                      \\
		\rule[0pt]{0pt}{10pt} 37 & Au (Au\_def)  &   0.000   &   0.019  &   0.000   &  0.015   &   \num{-0.020(40:45)}  &   -0.032  &                      \\
		\rule[0pt]{0pt}{10pt} 38 & Au (Au\_def)  &   0.000   &  -0.009  &   0.000   &  0.017   &   \num{-0.049(30:26)}  &   -0.043  &                      \\
		\rule[0pt]{0pt}{10pt} 39 & Au (Au\_def)  &   0.000   &   -0.022  &   0.000   &  0.006   &   \num{-0.011(33:30)}  &   -0.011  &                      \\
		\bottomrule
	\end{tabular}
	\caption{Structural parameters and vibrational amplitudes in \AA\ for the topmost four layers are derived from the LEED-IV analysis of the Au(111)-\rectcell-2(\tite). In addition to the LEED-IV results, the values from the DFT-relaxed slab are also provided. The atom number (N) is given according to the best-fit POSCAR. Only one atom of each group of symmetry-equivalent atom sites is shown in the table. Symmetry-equivalent atoms are marked by ''linking`` column in the best-fit POSCAR file. The values $\Delta x$, $\Delta y$ and $\Delta z$ represent deviations from the ideal bulk positions. The vibrational amplitudes, $\Delta u$, are the absolute best-fit values determined by LEED-IV. A vibrational amplitude is fitted for each atom label, noted in brackets. The vibrational amplitude for Au bulk atoms was set according to the known Debye temperature \cite{Kittel1980}.}
	\label{tab:rect}
\end{table}

\subsection{Monolayer \tite\ fit}
Table \ref{tab:single} presents the results of the structural analysis for the monolayer \tite\ film. As outlined in Section \ref{sm:methods}, the Au(111) substrate is excluded from the structural analysis. The $\Delta z$ values indicate deviations from the bulk \tite\ structure reported by \citet{Arnaud1981}. The in-plane lattice parameter is optimized to \qty{3.800(24)}{\AA}.
The optical potential $V_{0i}$ was set to \qty{4.75}{\eV} and the effective angle of incidence to $\Theta_{\mathrm{eff}} = \qty{0.4}{\degree}$, a well-established initial guess for the experimental setup used, as referenced in \cite{Kisslinger2020, Kisslinger2021, Kisslinger2023}. 
Both these values were not optimized. The statistical weights of the two domains, representing possible stacking sequences, were optimized to an $82:18$ ratio with an error of $\pm\qty{9}{\percent}$.
The total energy data basis amounts to $\Delta E=\qty{1602}{\eV}$. The fit's redundancy yields $\rho=\nicefrac{\Delta E}{4NV_{0i}}\approx8.4$, with a total of $N=10$ parameters used.

\begin{table}
	\begin{tabular}{lcc}
		\toprule
		atom (label) & $\Delta z$ & $\Delta u$ \\
		\midrule
		\rule[0pt]{0pt}{12pt} Te1 (Te\_top) &  \num{-0.052(19:18)} & \num{0.067(51:67)} \\
		\rule[0pt]{0pt}{12pt} Ti1 (Ti\_def) & \num{-0.018(17:19)} & \num{0.090(37:70)}  \\
		\rule[0pt]{0pt}{12pt} Te2 (Te\_bot) & 0.0 (ref.) & \num{0.065(50:65)}  \\
		\bottomrule
	\end{tabular}
	\caption{Structural parameters and vibrational amplitudes in \AA\ are derived from the LEED-IV analysis of the monolayer \tite\, excluding Au(111). For better comparison, the z values represent deviations from the published bulk \tite\ positions by \citet{Arnaud1981}, with the z position of the lower Te2 atom serving as a reference. The vibrational amplitudes, $\Delta u$, are the best-fit absolute values identified by LEED-IV. As indicated by the error of the vibrational amplitudes, there is no lower bound greater than zero where the error curve crosses $R+\mathrm{var}(R)$ again. Consequently, the lower bound of the provided uncertainty is the inverse of the absolute vibrational amplitude.}
	\label{tab:single}
\end{table}

\subsection{Multi-layer fit}
\label{sm:subsec_multi}
Table \ref{tab:multi} summarizes the results for the structural analysis of the multi-layer \tite\ film on Au(111). The $\Delta z$  values indicate deviations from the bulk \tite\ structure as published by \citet{Arnaud1981}.  The lowest of the four Te-Ti-Te trilayers was kept fixed, the z positions of the atoms allocated in the remaining layers were varied (9 structural parameters). The vibrational amplitudes of the two topmost Te atoms, as well as the topmost Ti atom, were optimized separately, while all other atoms are varied together in groups according to their type. In total, 9 structural and 5 vibrational amplitudes were fitted.
The in-plane lattice parameter is optimized to \qty{3.835(20)}{\AA}. 
The effective angle of incidence was optimized to $\Theta_{\mathrm{eff}} = \qty{0.54}{\degree}$ and the optimization of $V_{0i}$ yields \qty{5.8}{\eV}. Statistical weights of the two domains, representing possible stacking sequences, were optimized to an $82^{+8}_{-11}:18^{-8}_{+11}$ ratio.
The total energy data basis accumulates to $\Delta E=\qty{3520}{\eV}$, and a total of $N=19$ fit parameters are varied. The fit's redundancy is calculated as $\rho=\nicefrac{\Delta E}{4NV_{0i}}\approx8.0$.

\begin{table}
	\begin{tabular}{lcc}
		\toprule
		atom (label) & $\Delta z$ & $\Delta u$ \\
		\midrule
		\rule[0pt]{0pt}{12pt} Te1 (Te\_top) & \num{-0.280(14:14)} & \num{0.109(22:21)} \\
		\rule[0pt]{0pt}{12pt} Ti1 (Ti\_top) & \num{-0.236(18:19)} & \num{0.126(22:21)} \\
		\rule[0pt]{0pt}{12pt} Te2 (Te\_bot) & \num{-0.236(14:14)} & \num{0.069(22:34)} \\
		\rule[0pt]{0pt}{12pt} Te3 (Te\_def) & \num{-0.195(15:14)} & \num{0.069(38:39)} \\
		\rule[0pt]{0pt}{12pt} Ti2 (Ti\_def) & \num{-0.184(31:24)} & \num{0.075(48:70)} \\
		\rule[0pt]{0pt}{12pt} Te4 (Te\_def) & \num{-0.132(36:54)} & \num{0.069(38:39)} \\
		\rule[0pt]{0pt}{12pt} Te5 (Te\_def) & \num{-0.089(48:43)} & \num{0.069(38:39)} \\
		\rule[0pt]{0pt}{12pt} Ti3 (Ti\_def) & \num{-0.074(86:99)} & \num{0.075(48:70)} \\
		\rule[0pt]{0pt}{12pt} Te6 (Te\_def) & \num{-0.024(58:68)} & \num{0.069(38:39)} \\
		\rule[0pt]{0pt}{12pt} Te7 (Te\_def) & 0.000 (bulk) & \num{0.069(38:39)} \\ 
		\rule[0pt]{0pt}{12pt} Ti4 (Ti\_def) & 0.000 (bulk) & \num{0.069(38:39)} \\ 
		\rule[0pt]{0pt}{12pt} Te8 (Te\_def) & 0.000 (bulk) & \num{0.069(38:39)} \\ 
		\bottomrule
	\end{tabular}
	\caption{Structural parameters and vibrational amplitudes in \AA\ were determined from the LEED-IV analysis of the multi-layer \tite\ film on Au(111). The $\Delta z$ values represent deviations from the bulk positions. The lower Te atom of the third Te-Ti-Te trilayer (Te7) serves as the reference position for both the bulk and the fitted multi-layer structure. The vibrational amplitudes, $\Delta u$, are the absolute best-fit values identified by LEED-IV. For each atom label indicated in brackets, a specific vibrational amplitude is fitted.}
	\label{tab:multi}
\end{table}

\end{document}